\documentclass[twocolumn,aps,showpacs,prb,tightenlines,amsmath,amssymb,superscriptaddress,10pt]{revtex4-1}
\usepackage{graphicx}
\usepackage{amssymb}
\usepackage{dcolumn}
\usepackage{amsmath}
\usepackage{bm}
\usepackage{colordvi}

\begin{document}

\title{Current deformation and quantum inductance in mesoscopic capacitors}
\author{Y. Yin}
\thanks{Author to  whom correspondence should be addressed}
\email{yin80@scu.edu.cn.}
\affiliation{Laboratory of Mesoscopic and Low Dimensional Physics,
Department of Physics,
Sichuan University, Chengdu, Sichuan, 610064, China}
\date{\today}

\begin{abstract}

  We present a theoretical analysis of low frequency dynamics of a
  single-channel mesoscopic capacitor, which is composed by a quantum dot
  connected to an electron reservoir via a single quantum channel. At low
  frequencies, it is known that the Wigner-Smith delay time $\tau_W$ plays a
  dominant role and it can be interpreted as the time delay between the current
  leaving the dot and the current entering the dot. At higher frequencies, we
  find that another characteristic time $\tau_S$ can also be important. It
  describes the deformation of the leaving current to the entering one and hence
  can be referred as the deformation time. At sufficient low temperatures, the
  deformation time $\tau_S$ can be approximated from the second-order derivative
  of $\tau_W$ via a simple relation $\tau''_W/\tau^3_S=24/\hbar^2$. As the
  temperature increases, this relation breaks down and one has instead
  $\tau''_W/\tau^3_S \to 0$ in the high temperature limit. We further show that
  the deformation time $\tau_S$ can have a pronounced influence on the quantum
  inductance $L_q$ of the mesoscopic capacitor, leading to features different
  from the ones of the quantum capacitance. The most striking one is that $L_q$
  can change its sign as the temperature increases: It can go from positive
  values at low temperatures to large negative values at high temperatures. The
  above results demonstrate the importance of the deformation time $\tau_S$ on
  the ac conductance of the mesoscopic capacitor.

\end{abstract}

\pacs{73.23.-b, 
  72.10.-d,     
  72.21.La}     

\maketitle

\section{INTRODUCTION}

The understanding of the low-frequency ac conductance of quantum conductors has
attracted renewed interest in recent years.\cite{gabelli2006, gabelli2007,
  feve2007, bocquillon2013, dubois2013} In the linear response regime, it has
been demonstrated that the ac conductance is directly related to the
Wigner-Smith delay time of electrons.\cite{gabelli2006} This offers a means to
investigate the charge dynamics on a mesoscopic scale.\cite{gabelli2012} In the
nonlinear regime, the control and manipulation of a single electron have been
realized at gigahertz frequencies.\cite{feve2007,bocquillon2013, dubois2013}
This opens the way to the new generation devices which can serve as building
blocks for quantum electron optics and quantum information
processing.\cite{mahe2010, parmentier2012, bocquillon2012}

\begin{figure}
  \centering
  \includegraphics[width=7.5cm]{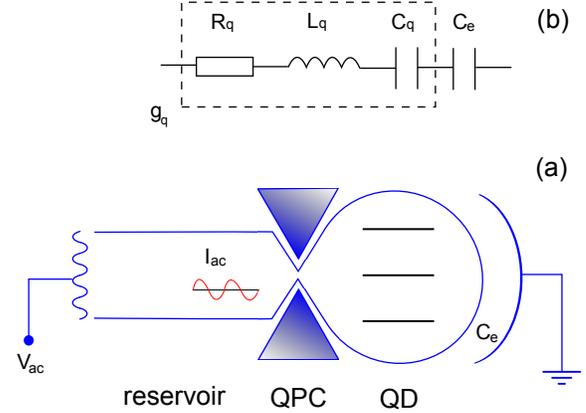}
  \caption{(Color online) (a) Mesoscopic capacitor. The tunneling between the QD
    and the reservoir are controlled via the QPC. An external ac field $V_{\rm
      ac}$ causes the carrier exchange between the QD and the reservoir, leading
    to an ac current $I_{\rm ac}$. The QD and the metallic electrode on the
    right are capacitively coupled, which does not permit carrier exchange. (b)
    The low-frequency equivalent circuit of the mesoscopic capacitor. The
    geometrical capacitance $C_e$ (formed between the QD and the right
    electrode) is in series with the quantum conductance $g_q$. At low
    frequencies, $g_q$ can be characterized by a quantum capacitor $C_q$, a
    charge relaxation resistance $R_q$ and a quantum inductance $L_q$.}
  \label{fig1}
\end{figure}

As an elementary structure in this area, the mesoscopic
capacitor\cite{buttiker1993, buttiker1993-1} plays a central role. It is
composed by a quantum dot (QD) and an electron reservoir, connected via a
quantum point contact (QPC), as illustrated in Fig.~\ref{fig1}. The QD is
capacitively coupled to a metallic electrode, which forms a geometrical
capacitor with capacitance $C_e$. The low-frequency ac conductance of the
mesoscopic capacitor has been extensively studied both
theoretically\cite{buttiker2007, nigg2006, nigg2008, moskalets2008, ringel2008,
  nigg2009, mora2010, hamamaoto2010, lee2011, filippone2012} and
experimentally.\cite{gabelli2006} It has been found that the Wigner-Smith delay
time $\tau_W$ is crucial to the ac conductance up to the second order of the
frequency $\omega$, leading to a quantum capacitance $C_q=e^2\tau_W/h$ and a
universal charge relaxation resistance $R_q=h/(2 e^2)$ for the single-channel
mesoscopic capacitor. The corresponding ac conductance $G(\omega)$ can be
expressed as
\begin{eqnarray}
  G(\omega) & = & -i \omega C_{\mu} (1 + i \omega R_q C_{\mu}) + O(\omega^2),
  \label{s1:eq1}
\end{eqnarray}
where $C_{\mu} = 1/(1/C_e+1/C_q)$ is usually referred as the electrochemical
capacitance.\cite{buttiker1993}

The effect of the Wigner-Smith delay time $\tau_W$ on the low-frequency ac
conductance can be vividly interpreted within the delayed current picture
introduced by Ringel {\em et al.}.\cite{ringel2008} They show that the total
current of the mesoscopic capacitor can be interpreted in terms of an incoming
current and an outgoing current. While the incoming current responds
instantaneously to the external driving field, the outgoing current is delayed
by $\tau_W$ with respect to the incoming one. Such picture well captures the
behavior of ac conductance at low frequencies. It is worth noting that at higher
frequencies, the effect of the Wigner-Smith delay time are expected to be more
pronounced. Wang {\em et al.} have shown that up to the third order of the
frequency, it can lead to a quantum inductance $L_q=R_q \tau_W/12$ under the
resonance condition when the QD levels are aligned with the Fermi energy of the
reservoir.\cite{wang2007} The corresponding $G(\omega)$ can be written as
\begin{eqnarray}
  \hspace{-0.5cm}G(\omega) & = & -i \omega C_{\mu} (1 + i \omega R_q C_{\mu} - \omega^2
  C^2_{\mu} R^2_q + \omega^2 C_{\mu} L_q ) \nonumber\\
  &&  + O(\omega^3).
  \label{s1:eq2}
\end{eqnarray}
The above result highlights the importance of the Wigner-Smith delay time on the
charge dynamics of the mesoscopic capacitor.

One may wonder whether there are other characteristic times that can also play a
role on the ac conductance, especially at higher frequencies. This is the main
motivation of this work. To study this question, we first reexamine the delayed
current picture and find that the current delay characterized by the
Wigner-Smith delay time $\tau_W$ can only describe the behavior of the ac
conductance $G(\omega)$ up to the second order of the frequency. As the
frequency goes higher, another effect ---current deformation--- can be
important. A new characteristic time ---deformation time $\tau_S$--- is then
introduced to describe such effect. While the Wigner-Smith delay time $\tau_W$
is decided by the density of states of the mesoscopic capacitor, the deformation
time $\tau_S$ is related to its second-order derivative. The two characteristic
times can be related via a simple relation
\begin{eqnarray}
\frac{\tau^{''}_W}{\tau^3_S} & = & \frac{24}{\hbar^2}. \label{s1:eq3}
\end{eqnarray}
By incorporating the current deformation into the delayed current picture(which
can be referred as delayed-deformed picture), we find that $\tau_S$ can manifest
itself in the quantum inductance $L_q$ as
\begin{eqnarray}
  L_q & = & R_q \tau_W [ 4\big( \frac{\tau_S}{\tau_W} \big)^3 + \frac{1}{6} ].
  \label{s1:eq0}
\end{eqnarray}
Due to the effect of $\tau_S$, the quantum inductance $L_q$ can have quite
different behaviors from the quantum capacitance $C_q$. Specifically, $L_q$ can
exhibit dips around the resonances where $C_q$ exhibits peaks. The $L_q$
obtained by Wang {\em et al.} in Ref.~\onlinecite{wang2007} can be treated as a
specific case of Eq.~\eqref{s1:eq0} at the resonances.

We further validate the conclusions obtained from the delayed-deformed picture
by performing more realistic calculations within the non-equilibrium Green's
function (NEGF) formalism, where the effect of the charging energy in the QD and
the nonzero temperature have been taken into consideration. We find that the
relation between the deformation time $\tau_S$ and the Wigner-Smith delay time
$\tau_W$ [Eq.~\eqref{s1:eq3}] is a good approximation at sufficient low
temperatures despite the presence of the charging energy. As the temperature
increases, such relation tends to break down and one has instead
\begin{eqnarray}
\frac{\tau^{''}_W}{\tau^3_S} & \to & 0, \label{s1:eq4}
\end{eqnarray}
in the high temperature limit. We also find that the deformation time $\tau_S$ do
have a pronounced impact on the quantum inductance $L_q$. It can not only lead
to dips around the resonances, but also make $L_q$ changes its sign at nonzero
temperatures: $L_q$ can go from positive values at low temperatures to large
negative values at high temperatures. Thus, just like the universality of the
charge relaxation resistance $R_q$, the positive definiteness of $L_q$ can also
be regarded as a signature of the quantum coherent transport. The above results
demonstrate the importance of the deformation time $\tau_S$ on the ac
conductance of the mesoscopic capacitor.

The paper is organized as follows: In Sec.~\ref{sec2}, we generalize the delayed
current picture to include the current deformation effect. In Sec.~\ref{sec3},
we present the Hamiltonian and the NEGF formalism. The numerical results from
the NEGF formalism are discussed in Sec.~\ref{sec4}. We summarized in
Sec.~\ref{sec5}.

\section{DELAYED-DEFORMED CURRENT}
\label{sec2}

In this section, we generalize the delayed current picture to include the
current deformation effect.

Following the scattering formalism,\cite{blanter2000} the current operator for
the single-channel mesoscopic capacitor can be decomposed into an incoming part
$\hat{I}_{+}$ and outgoing part $\hat{I}_{-}$, which can be expressed as
\begin{eqnarray}
  \hat{I}(t) & = & \hat{I}_{+}(t) - \hat{I}_{-}(t), \label{s3:eq1-1}\\
  \hat{I}_{+}(t) & = & \frac{e}{2 \pi \hbar} \int dE \hat{n}_{+}(E, t), \label{s3:eq1-2}\\
  \hat{I}_{-}(t) & = & \frac{e}{2 \pi \hbar} \int dE \hat{n}_{-}(E, t), \label{s3:eq1-3}
\end{eqnarray}
where $\hat{n}_{+}$($\hat{n}_{-}$) is the operator describing the occupation
numbers of the incoming(outgoing) channel. They can be written as
\begin{eqnarray}
  \hat{n}_{+}(E, t) & = & \int d\omega e^{-i \omega t} \hat{a}^{\dagger} (E-\hbar\omega/2) \hat{a}
  (E+\hbar\omega/2), \label{s3:eq2-1}\\
  \hat{n}_{-}(E, t) & = & \int d\omega e^{-i \omega t} \hat{b}^{\dagger} (E-\hbar\omega/2) \hat{b}
  (E+\hbar\omega/2), \label{s3:eq2-2}
\end{eqnarray}
where $\hat{a}(E)$[$\hat{b}(E)$] and
$\hat{a}^{\dagger}(E)$[$\hat{b}^{\dagger}(E)$] are the creation and annihilation
operators of electrons in the incoming(outgoing) channel with energy $E$,
respectively.

For the case of elastic scattering, the operators $\hat{a}$ and $\hat{b}$ are
related via the scattering matrix, which can be described by just a pure phase
factor $\phi$ for the single-channel system,\cite{pretre1996, buttiker1992,
  levinson2000}
\begin{eqnarray}
  \hat{b}(E) & = & S(E) \hat{a}(E), \label{s3:eq3-1}\\
  S(E) & = & e^{i \phi(E)}. \label{s3:eq3-2}
\end{eqnarray}
By substituting Eqs.~\eqref{s3:eq3-1} and \eqref{s3:eq3-2} into
Eqs.~\eqref{s3:eq2-1} and \eqref{s3:eq2-2}, one obtains the relation between the
occupation number operator of incoming and outgoing electrons,
\begin{eqnarray}
  \hspace{-0.5cm}\hat{n}_{-}(E, t) & = & \int dt' \hat{n}_{+}(E, t') A(E, t - t'), \label{s3:eq4-1}
\end{eqnarray}
where the effects of the scattering are attributed to the integral kernel
$A(E,t)$. It can be expressed in terms of the scattering phase factor as
\begin{eqnarray}
  \hspace{-0.5cm} A(E, t) & = & \int \frac{d\omega}{2 \pi} e^{-i \omega t} e^{-i [\phi(E-\hbar\omega/2) -
    \phi(E + \hbar\omega/2)] }. \label{s3:eq4-2}
\end{eqnarray}

If the scattering phase factor $\phi(E)$ is slow-varying with respect to the
energy $E$, the integral kernel $A(E, t)$ can be expanded with respect to the
frequency $\omega$, yielding the low-frequency expansion
\begin{eqnarray}
  A(E, t) & = & \int \frac{d\omega}{2 \pi} e^{-i \omega t} e^{i \omega \tau_W +
    i \omega^3 \tau^3_S + O(\omega^3) }. \label{s3:eq5}
\end{eqnarray}
The two parameters $\tau_W$ and $\tau_S$ in Eq.~\eqref{s3:eq5} can be written as
\begin{eqnarray}
  \tau_W(E) & = & 2 \pi \hbar \rho(E), \label{s3:eq6-1} \\
  \tau_S(E) & = & \frac{\hbar}{2} \sqrt[3]{\frac{2 \pi \rho''(E)}{3}}, \label{s3:eq6-2}
\end{eqnarray}
where $\rho(E)=\rm{Tr}[S^{\dagger}\partial_E S]/(2 \pi i)$ representing the
density of states for the capacitor plate,\cite{nigg2006} while $\rho{''}(E)$
denoting the second-order derivative of $\rho(E)$ with respect to the energy
$E$.

Equation \eqref{s3:eq5} indicates that at low frequencies, the effect of the
scattering can be described by the two parameters $\tau_W$ and $\tau_S$. The
parameter $\tau_W$ is just the Wigner-Smith delay time, indicating that the due
to the scattering, the outgoing current is delayed from the incoming current by
$\tau_W$, while the profile of the outgoing current remains the same as the
incoming one. The parameter $\tau_S$, which also has dimension of time,
indicates that due to the scattering, the profile of the outgoing current is
deformed from the incoming one. The magnitude of the deformation can be
quantitatively described by $\tau_S$, hence it can be referred as "deformation"
time. It is worth emphasizing that according to Eqs.~\eqref{s3:eq6-1} and
\eqref{s3:eq6-2}, the deformation time $\tau_S$ can be related to the
Wigner-Smith delay time $\tau_W$ as
\begin{eqnarray}
  \frac{\tau^{''}_W}{\tau^3_S} & = & \frac{24}{\hbar^2}. \label{s3:eq6-3}
\end{eqnarray}

Both $\tau_W$ and $\tau_S$ can manifest itself in the quantum conductance $g_q$
of the mesoscopic capacitor[Fig.~\ref{fig1}(b)]. Up to the first-order of the
frequency $\omega$, only the effect of the current delay can play a role. The
corresponding quantum conductance can be approximated as
\begin{eqnarray}
  g_q(\omega) & \approx & \frac{e^2}{h} (1 - e^{i \omega \tau_W(E_F) }), \label{s3:eq7-1}
\end{eqnarray}
with $E_F$ being the Fermi energy of the reservoir. This is just the result
obtained within the delayed current picture.\cite{ringel2008} Up to the
second-order of the frequency $\omega$, the effect of the current deformation
can also be important. The quantum conductance $g_q$ including this effect
becomes (see Appendix~\ref{app} for derivation)
\begin{eqnarray}
  g_q(\omega) & \approx & \frac{e^2}{h} [1 - 2 e^{i \omega \tau_W(E_F) } e^{i \omega^3
    \tau^3_S(E_F)} \nonumber\\
  &&\hspace{2.5cm} + e^{i \omega \tau_W(E_F) }]. \label{s3:eq7-2}
\end{eqnarray}
Equation~\eqref{s3:eq7-2} demonstrates the effect of the current deformation on
the charge dynamics of the mesoscopic capacitor.

As the deformation time $\tau_S$ does not play a role on the quantum
conductance $g_q$ up to the first-order of the frequency, it can not affect the
quantum capacitance $C_q$ and the relaxation resistance $R_q$ of the mesoscopic
capacitor. However, it does have a non-negligible influence on the quantum
inductance $L_q$. To show this, we calculate $C_q$, $R_q$ and $L_q$ by matching
the impedance of the mesoscopic capacitor\cite{gabelli2006, buttiker1993,
  wang2007}
\begin{eqnarray}
  Z(\omega) & = & \frac{1}{g_q(\omega)} - \frac{1}{i \omega C_e},
  \label{s3:eq8}
\end{eqnarray}
to the corresponding formula for a classical RLC circuits
\begin{eqnarray}
  Z(\omega) & = & \frac{i}{\omega} \Big( \frac{C_q C_e}{C_q + C_e} - i \omega R_q -
  \omega^2 L_q \Big),
  \label{s3:eq9}
\end{eqnarray}
where the quantum conductance $g_q$ is given by Eq.~\eqref{s3:eq7-2}. By
comparing Eq.~\eqref{s3:eq8} to Eq.~\eqref{s3:eq9}, one obtains
\begin{eqnarray}
  C_q & = & \frac{e^2}{h} \tau_W, \label{s3:eq10-1}\\
  R_q & = & \frac{h}{2 e^2}, \label{s3:eq10-2}\\
  L_q & = & R_q \tau_W [ 4\big( \frac{\tau_S}{\tau_W} \big)^3 + \frac{1}{6} ]. \label{s3:eq10-3}
\end{eqnarray}
The influence of the deformation time $\tau_S$ on the quantum inductance $L_q$ can
be clearly seen from Eq.~\eqref{s3:eq10-3}.

\begin{figure}
  \centering
  \includegraphics[width=8.5cm]{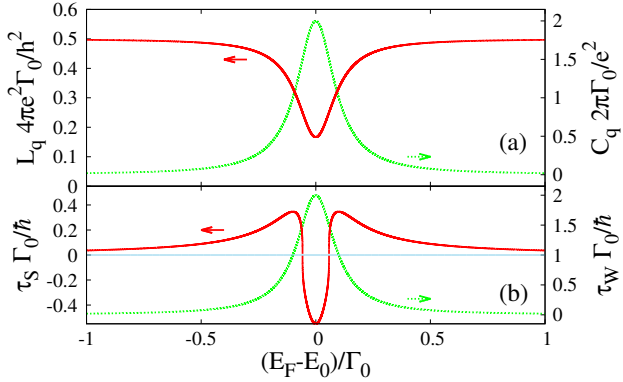}
  \caption{(Color online) (a) The quantum inductance $L_q$ (red solid curve) and
    quantum capacitance $C_q$ (green dotted curve) as function of the Fermi
    energy $E_F$. (b) The corresponding deformation time $\tau_S$ (red solid
    curve) and the Wigner-Smith delay time $\tau_W$ (green dotted curve) as
    function of the Fermi energy $E_F$. The thin skyblue line indicates the zero
    value for $\tau_S$.}
  \label{fig2}
\end{figure}

Due to the additional contribution from the deformation time $\tau_S$, the
quantum inductance $L_q$ can have quite different behaviors from the ones of the
quantum capacitance $C_q$. To illustrate this, we suppose the density of states
$\rho(E)$ of the mesoscopic capacitor is given by a single Lorentzian peak
around $E_0$ with width $\Gamma_0$,
\begin{eqnarray}
  \rho(E) & = & \frac{1}{\pi} \frac{\Gamma_0}{ (E-E_0)^2 + \Gamma^2_0 }, \label{s3:eq11}
\end{eqnarray}
then according to Eqs.~\eqref{s3:eq10-1} and \eqref{s3:eq10-3}, $C_q$ also
exhibits a Lorentzian peak as a function of $E_F$, while $L_q$ exhibits a dip
around $E_0$, as illustrated in Fig.~\ref{fig2}(a). At the resonance ($E_F=E_0$)
where $C_q$ reaches its maximum, $L_q$ reaches its minimum value $L^{\rm min}_q
= R_q \tau^{\rm max}_W/12$ with $\tau^{\rm max}_W=2\hbar/\Gamma_0$. This is just
the result obtained by Wang {\em et al.}  in Ref.~\onlinecite{wang2007}. Far
from the resonance ($|E_F-E_0| \gg \Gamma_0$) where $C_q$ tends to zero, $L_q$
reaches its maximum value $L^{\rm max}_q = 3 L^{\rm min}_q$. By comparing to the
corresponding $\tau_S$ and $\tau_W$ in Fig.~\ref{fig2}(b), one can see that the
dip in $L_q$ can be attributed to the contribution of $\tau_S$, which also
exhibits a dip around the resonance.

It is worth noting that the relation between $L_q$ and $\tau_S$ can offer a way
to detect the detailed structure of the density of states $\rho(E)$, since
$\tau_S$ is related to the second-order derivative of $\rho(E)$
[Eq.~\eqref{s3:eq6-2}]. It is interesting to remind that the first-order
derivative of $\rho(E)$ can be accessed via the thermoelectric
capacitance.\cite{lim2013} This suggests that by combining the charge and
thermoelectric admittance, one can obtain more complete information of
mesoscopic systems.

To summarize this section, we have generalize the delayed current picture to
include the current deformation effect into consideration. Such effect can be
quantitatively described by the deformation time $\tau_S$, which is related to
the Wigner-Smith delay time via a simple relation Eq.~\eqref{s3:eq6-3}. The
deformation time $\tau_S$ can have a pronounced impact on the quantum inductance
$L_q$ of the mesoscopic capacitor, making $L_q$ having quite different behaviors
from the ones of the quantum capacitance $C_q$.

\section{NEGF FORMALISM}
\label{sec3}

Although the delayed-deformed current picture offers a vivid interpretation of
the charge dynamics of the mesoscopic capacitor, the approximation used in the
derivation is rather crude. Some effects, such as the breakdown of the
universality of $R_q$,\cite{nigg2006} are ignored in such picture. To further
validate the conclusions from the delayed-deformed current picture, we perform
more realistic calculations within the framework of non-equilibrium Green's
function (NEGF) formalism.

Let us first present the Hamiltonian of the mesoscopic capacitor which is
illustrated in Fig.~\ref{fig1}. It can be written as\cite{ringel2008}
\begin{eqnarray}
  H & = & H_L + H_D + H_{\rm LD},
  \label{s2:eq1}
\end{eqnarray}
where $H_L$, $H_D$ and $H_{\rm LD}$ describe the reservoir, the QD and
their coupling, respectively. The reservoir Hamiltonian $H_L$ is derived from a
one-dimensional tight-binding model, which can be written as
\begin{eqnarray}
  H_L & = & \int dk \varepsilon(k) a^{\dagger}_k a_k,
  \label{s2:eq2}
\end{eqnarray}
where $\varepsilon(k) = - 2 t_0 \cos(k)$ is the dispersive relation with $t_0$
being the hopping between adjacent sites. The Hamiltonian of the QD, including
the single-particle part and interactions, can be expressed as
\begin{eqnarray}
  H_D & = & \sum^{n_d}_{n=1} \epsilon_n d^{\dagger}_n d_n + \frac{E_C}{2} (\hat{N} - \frac{n_d}{2})^2,
  \label{s2:eq3}
\end{eqnarray}
where $\epsilon_n=n \Delta$ with $\Delta$ being the level spacing. $E_C=e^2/C_e$
describes the charging energy with $C_e$ being the geometrical
capacitance. $\hat{N}=\sum^{n_d}_{n=1} d^{\dagger}_n d_n$ is the number operator
of the electrons in the QD with $n_d$ devoting the number of QD levels. The
coupling $H_{LD}$ can be written as
\begin{eqnarray}
  H_{LD} & = & \sum_n \int dk ( t_{\rm kn} d^{\dagger}_n a_k + {\rm H.c.}),
  \label{s2:eq4}
\end{eqnarray}
with $t_{\rm kn}$ being the coupling matrix element.

Within the NEGF formalism, the quantum admittance $g_q(\omega)$ can be
calculated in the wide-band-limit\cite{haugbook, bgwang1999, ma1999} as
\begin{eqnarray}
  g_q(\omega) & = & - i \frac{e^2}{h} \int d\omega' {\rm Tr}[
  G^r_D(\omega+\omega')\frac{\Gamma}{\hbar}G^a_D(\omega') ] \nonumber\\
  && \times [f(\omega') - f(\omega+\omega')],
  \label{s2:eq5}
\end{eqnarray}
where $G^{\rm r/a}_D(\omega)$ represents the equilibrium retarded/advanced Green
function of the QD while $\Gamma$ describes the level-width function of the QD
due to the coupling to the reservoir.\cite{haugbook}
$f(\omega)=1/[1+\exp(\beta(\hbar\omega-E_F))]$ represents the equilibrium
electron distribution, with $E_F$ being the Fermi level and $\beta=1/(k_BT)$
being the inverse temperature. The Taylor expansion of $g_q(\omega)$ with
respect to the frequency $\omega$ reads
\begin{eqnarray}
  \hspace{-1cm} g_q(\omega) & = & e_0 + e_1 \omega + e_2 \omega^2 + O(\omega^3), \label{s2:eq6-1}\\
  \hspace{-1cm} e_0 & = & - \int d\omega  f'(\omega) {\rm Tr} [
  G^r_D(\omega)\frac{\Gamma}{\hbar}G^a_D(\omega) ], \label{s2:eq6-2}\\
  \hspace{-0.0cm} e_1 & = & - \int d\omega \frac{f'(\omega)}{2} {\rm Tr} [
  (G^r_D(\omega))'\frac{\Gamma}{\hbar}G^a_D(\omega) \nonumber\\
  && \hspace{-0.4cm} - G^r_D(\omega)\frac{\Gamma}{\hbar}(G^a_D(\omega))' ], \label{s2:eq6-3} \\
  \hspace{-1cm} e_2 & = & - \int d\omega  \frac{f'(\omega)}{6} {\rm Tr} [
  (G^r_D(\omega))''\frac{\Gamma}{\hbar}G^a_D(\omega) \nonumber\\
  && \hspace{-0.4cm} + G^r_D(\omega)\frac{\Gamma}{\hbar}(G^a_D(\omega))'' -
  (G^r_D(\omega))'\frac{\Gamma}{\hbar}(G^a_D(\omega))' ], \label{s2:eq6-4}
\end{eqnarray}
where $(G^{r/a}_D(\omega))'$ and $(G^{r/a}_D(\omega))''$ represent the
first-order and second-order derivatives of the retarded/advanced QD Green
function with respect to the frequency $\omega$.

By substituting Eqs.~(\ref{s2:eq6-1}-\ref{s2:eq6-4}) into Eq.~(\ref{s3:eq8}) and
comparing to Eq.~(\ref{s3:eq9}-\ref{s3:eq10-3}), one obtains the quantum
capacitor $C_q$, charge relaxation resistance $R_q$ and the quantum inductance
$L_q$ as
\begin{eqnarray}
  C_q & = & \frac{e^2}{h} e_0, \label{s2:eq8-1}\\
  R_q & = & \frac{\tau_W}{2 C_q}, \label{s2:eq8-2}\\
  L_q & = & R_q \tau_W [ 4\big( \frac{\tau_S}{\tau_W} \big)^3 + \frac{1}{6} ] \label{s2:eq8-3},
\end{eqnarray}
where the two characteristic times $\tau_W$ and $\tau_S$ can be expressed as
\begin{eqnarray}
  \tau_W & = & -2i\frac{e_1}{e_0}, \label{s2:eq9-1}\\
  \tau_S & = & \sqrt[3]{\frac{5}{24} - \frac{e_0 e_2}{4 e^2_1}}. \label{s2:eq9-2}
\end{eqnarray}

To obtain $C_q$, $R_q$ and $L_q$, one needs to find the equilibrium
retarded/advanced QD Green function $G^{\rm r/a}_D$. They can be calculated
self-consistently within the Hartree-Fock approximation as\cite{ringel2008,
  nigg2006}
\begin{eqnarray}
  G^{r}_D & = & [ \hbar \omega - H + \frac{i \gamma}{2} ]^{-1}, \label{s2:eq10-1}\\
  H_{\rm mn} & = & \delta_{\rm mn} d + E_C\big( \delta_{\rm nm}
  \sum_{\bar{n}} Q_{\rm \bar{n}\bar{n}} - Q_{\rm nm} \big), \label{s2:eq10-2}\\
  Q_{\rm mn} & = & - \hbar \int \frac{d\omega}{\pi} f(\omega) {\rm
    Im}[G^{r}_D]_{\rm mn}. \label{s2:eq10-3}
\end{eqnarray}
In the calculation, we have assume all the QD levels coupled to the lead with
the same strength, i.e., $\Gamma_{\rm mn} = \gamma$.\cite{buttiker2007,
  ringel2008} Following Refs.~\onlinecite{buttiker2007, brouwer1997}, we choose
the coupling $\gamma$ as
\begin{eqnarray}
\gamma & = & \frac{\Delta}{\pi D}\Big( 2 - D - 2\sqrt{1-D} \Big),
\label{s2:eq11}
\end{eqnarray}
where $D$ describes the probability for transmission through the QPC. It can be
related to the Fermi energy $E_F$ in the lead as\cite{buttiker1990}
\begin{eqnarray}
D & = & \frac{1}{1 + e^{- a E_F/\Delta}},
\label{s2:eq12}
\end{eqnarray}
with $a$ being a constant depends on the detail structure of the QPC potential.

\section{NUMERICAL RESULTS}
\label{sec4}

The computations in this section are performed for the QD with $29$ levels.  The
parameter $a$ of the QPC is chosen to be $1.9$. The Fermi energy $E_F$ and the
QD charging energy $E_C$ are all measured in units of the QD level spacing
$\Delta$.

\begin{figure}
  \centering
  \includegraphics[width=8.5cm]{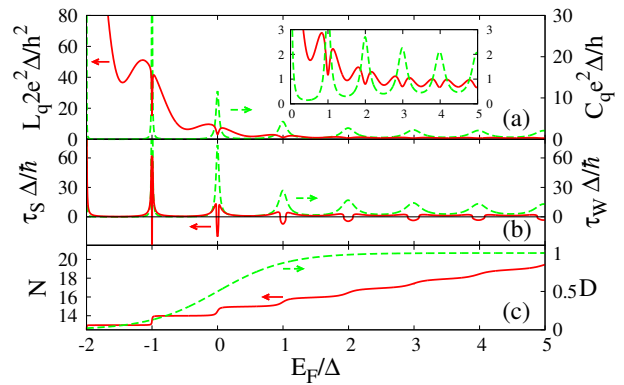}
  \caption{(Color online) (a) The quantum inductance $L_q$ (red solid curve) and
    quantum capacitance $C_q$ (green dashed curve) as function of the Fermi
    energy $E_F$. The zoom of $L_q$ and $C_q$ in the region $E_F/\Delta \in [0,
    5]$ are plotted in the inset. (b) The deformation time $\tau_S$ (red solid
    curve) and the Wigner-Smith delay time $\tau_W$ (green dashed curve) as
    function of the Fermi energy $E_F$. The thin black line indicates the zero
    value. (c) The total dot charge $N$ of the QD (red solid curve) and the
    transmission probability $D$ through the QPC (green dashed curve) as
    function of the Fermi energy. In all the figures, the temperature is set to
    $0$~K and the charging energy $E_C$ is set to $0$.}
  \label{fig3}
\end{figure}

We start the discussion from the simplest case where the temperature $T=0$~K and
the charging energy $E_C=0$.\cite{comment1} Let us first compare the behaviors
of the quantum inductance $L_q$ and the quantum capacitance $C_q$.  The $L_q$
and $C_q$ as function of the Fermi energy $E_F$ are plotted in
Fig.~\ref{fig3}(a).  The corresponding Wigner-Smith delay time $\tau_W$ and
deformation time $\tau_S$ are plotted in Fig.~\ref{fig3}(b). We also plot the
total dot charge $N$ and the probability for transmission through the QPC $D$ in
Fig.~\ref{fig3}(c) for comparison. From the figure, one can see that although
both $L_q$ and $C_q$ exhibit distinct oscillations as $E_F$ varies, the detail
structure of these oscillations are different. For small $E_F$ where the QPC is
close to pinch-off ($D \ll 1$), $C_q$ exhibits single sharp peaks at the
resonances where the transfer of an electron into the QD is permitted. For large
$E_F$ where the QPC is opened ($D \rightarrow 1$), the peak is broadened and its
height is decreased. On the contrary, $L_q$ exhibits sharp dips at the
resonances when the QPC is close to pinch-off, with two shoulder peaks appearing
at both sides of the dip. As the QPC is opened, such dip-double-peak structures
are suppressed into smooth shallow valleys.

By comparing to the corresponding $\tau_S$ and $\tau_W$, one can see that the
behavior of $C_q$ is solely decided by $\tau_W$, while the dip-double-peak
structures in $L_q$ can be attributed to the contribution from $\tau_S$. Hence
one concludes that the deformation time $\tau_S$ does play an important role on
the quantum inductance $L_q$. It can make $L_q$ exhibiting dips around the
resonances. These conclusions agree with the interpretation of the
delayed-deformed current picture Given in Sec.~\ref{sec2}.

\begin{figure}
  \centering
  \includegraphics[width=8.0cm]{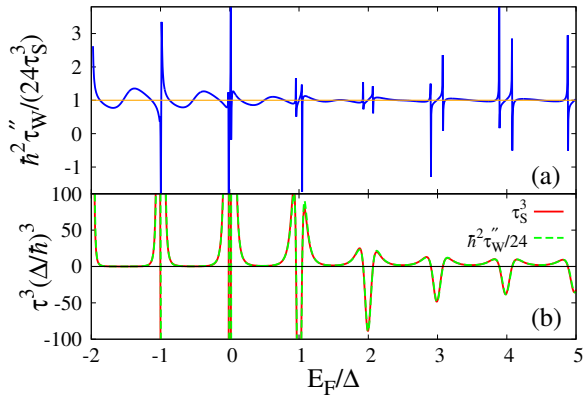}
  \caption{(Color online) (a) The ratio $\hbar^2\tau^{''}_W/(24\tau^3_S)$ as a
    function of $E_F$. The thin orange line indicates the value $1$. (b) The
    quantities $\tau^3_S$ (red solid curve) and $\hbar^2\tau^{''}_W/24$ (green
    dashed curve) as function of $E_F$. They are measured in units of
    $(\Delta/\hbar)^3$. The thin black line indicates the value $0$. In both
    figures, the temperature is set to $0$~K and the charging energy $E_C$ is
    set to $0$.}
  \label{fig4}
\end{figure}

Now let us discuss the relation between $\tau_W$ and $\tau_S$. The
delayed-deformed current picture predicts that they can be related via
Eq.~\eqref{s3:eq6-3}. To check this, we plot the ratio
$\hbar^2\tau^{''}_W/(24\tau^3_S)$ as a function of $E_F$ in
Fig.~\ref{fig4}(a). From the figure, one can see that the ratio
$\hbar^2\tau^{''}_W/(24\tau^3_S)$ is not exactly equal but quite close to the
value $1$. Relative large deviations occur only in the vicinity of the
resonances. However, the deviations are modest and the two quantities $\tau^3_S$
and $\hbar^2\tau^{''}_W/24$ agree quite well, as can be seen from
Fig.~\ref{fig4}(b). This indicates that although the relation
Eq.~\eqref{s3:eq6-3} derived from the delayed-deformed picture is not exact, it
can be regarded as a good approximation.

\begin{figure}
  \centering
  \includegraphics[width=9cm]{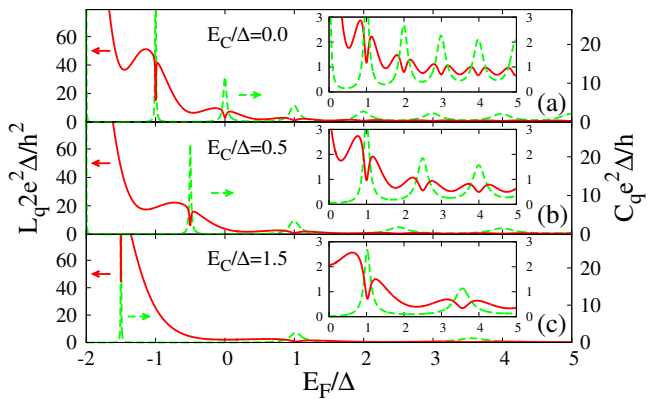}
  \caption{(Color online) The quantum inductance $L_q$ and quantum capacitance
    $C_q$ as function of the Fermi energy $E_F$ for $E_C/\Delta=0.0$ (a), $E_C/\Delta=0.5$ (b)
    and $E_C/\Delta=1.5$ (c). The zooms of $L_q$ and $C_q$ in the region $E_F/\Delta
    \in [0, 5]$ are plotted in the insets. In all the figures, the temperature
    is set to $0$~K.}
  \label{fig5}
\end{figure}

Next we turn to study the effect of the charging energy $E_C$. In
Fig.~\ref{fig5}, we plot the zero-temperature $L_q$ and $C_q$ as function of the
Fermi energy $E_F$ with different charging energy $E_C$. From the figure, one
can still identify the dip-double-peak structures in $L_q$ around the
resonances, even for nonzero charging energy $E_C$. Note that as $E_C$
increases, the dip-double-peak structures tend to be smeared out, making the
oscillations in $L_q$ less pronounced. This is similar to the suppression of
oscillations in $C_q$, which has been reported in previous
works.\cite{ringel2008, matveev1995}

\begin{figure}
  \centering
  \includegraphics[width=9cm]{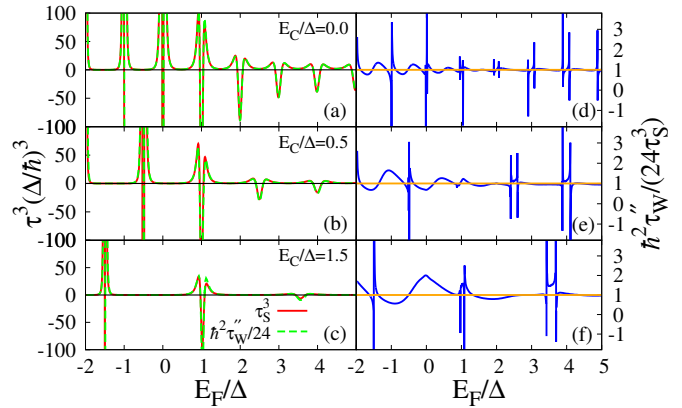}
  \caption{(Color online) The quantities $\tau^3_S$ (red solid curve) and
    $\hbar^2\tau^{''}_W/24$ (green dashed curve) as function of $E_F$ for
    $E_C/\Delta=0.0$ (a), $E_C/\Delta=0.5$ (b) and $E_C/\Delta=1.5$ (c). They
    are measured in units of $(\Delta/\hbar)^3$. The thin black lines indicate
    the value $0$. The corresponding ratios $\hbar^2\tau^{''}_W/(24\tau^3_S)$
    are plotted in (d), (e) and (f), respectively. The thin orange lines
    indicate the value $1$. In all the figures, the temperature is set to
    $0$~K.}
  \label{fig6}
\end{figure}

The corresponding ratio $\hbar^2\tau^{''}_W/(24\tau^3_S)$ are plotted in
Fig.~\ref{fig6}(d-f). One can see that for large $E_F$ when the QPC is opened,
the ratio is still quite close to $1$ and is not sensitive to $E_C$. For small
$E_F$ when the QPC is close to pinch-off, the ratio is relatively sensitive and
it can deviate from the value $1$ as $E_C$ increases. However, the deviation is
still small since the two quantities $\tau^3_S$ and $\hbar^2\tau^{''}_W/24$
agree quite well, as can be seen from Fig.~\ref{fig6}(a-c). This indicates that
the relation between $\tau_W$ and $\tau_S$ given by Eq.~\eqref{s3:eq6-3} is
still a good approximation for nonzero charging energy $E_C$, especially for the
cases with large $E_F$ when the QPC is opened.

The previous results justify the conclusions obtained from the delayed-deformed
picture: (1) The deformation time $\tau_S$ can play an important role on the
quantum inductance $L_q$, leading to dips around the resonances. (2) The
deformation time $\tau_S$ can be approximated from the Wigner-Smith delay
time $\tau_W$ via the simple relation Eq.~\eqref{s3:eq6-3}. These conclusions
hold at zero temperature, despite the presence of the charging energy.

\begin{figure}
  \centering
  \includegraphics[width=9cm]{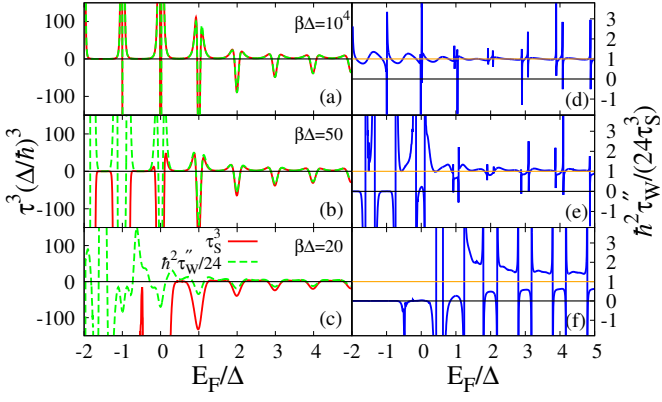}
  \caption{(Color online) The quantities $\tau^3_S$ (red solid curve) and
    $\hbar^2\tau^{''}_W/24$ (green dashed curve) as function of $E_F$ at inverse
    temperature $\beta \Delta = 10^4$ (a), $\beta \Delta = 50$ (b) and $\beta
    \Delta = 20$ (c). The thin black lines in (a-c) indicate the zero value. The
    corresponding ratios $\hbar^2\tau^{''}_W/(24\tau^3_S)$ are plotted in (d),
    (e) and (f), respectively. The thin black lines in (d-f) indicate the zero
    value, while the thin orange lines represent the value $1$. In all the
    figures, the charging energy $E_C$ is set to $0$.}
  \label{fig7}
\end{figure}

It is then natural to ask what happens to the quantum inductance $L_q$ and
deformation time $\tau_S$ at nonzero temperatures. To study this, we first
concentrate on the quantities $\tau^3_S$ and $\hbar^2\tau^{''}_W/24$ as function
of $E_F$ at different temperatures without the charging energy $E_C$ in
Fig.~\ref{fig7}(a-c). The corresponding ratio $\hbar^2\tau^{''}_W/(24\tau^3_S)$
are also plotted in Fig.~\ref{fig7}(d-f). From the figure, one can see that at
high temperatures, the quantity $\tau^3_S$ (red solid curve) disagrees with
$\hbar^2\tau^{''}_W/24$ (green dashed curve). Accordingly, the corresponding
ratio $\hbar^2\tau^{''}_W/(24\tau^3_S)$ tends to go from the value $1$ at low
temperatures to the value $0$ at high temperatures. Such effect is more
pronounced in the small $E_F$ region where the QPC is close to pinch-off. It is
worth noting that the increasing of the temperature can induce an overall
decreasing of the quantity $\tau^3_S$, making $\tau^3_S$ become large negative
values at high temperatures.

\begin{figure}
  \centering
  \includegraphics[width=8.5cm]{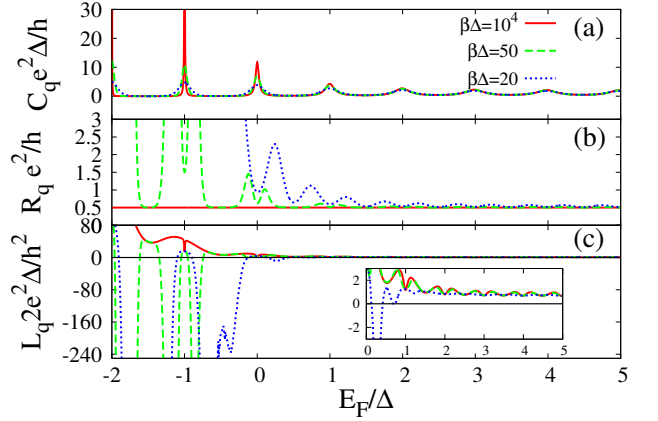}
  \caption{(Color online) The quantum capacitance $C_q$ (a) , charge relation
    resistance $R_q$ (b) and quantum inductance $L_q$ (c) as function of Fermi
    energy $E_F$ at different inverse temperatures. The zooms of $L_q$ in the
    region $E_F/\Delta \in [0, 5]$ are plotted in the insets of (c). The thin
    black lines in (c) and the inset indicate the zero value. In all the figures,
    the charging energy $E_C$ is set to $0$.}
  \label{fig8}
\end{figure}

The large negative $\tau_S$ can have a pronounced influence on the quantum
inductance $L_q$, leading to quite different behaviors from the ones in the zero
temperature limit. This is illustrated in Fig.~\ref{fig8}. From the figure, one
can see that $L_q$ can go from positive to negative in the region
$E_F/\Delta<0.73$ as the temperature increases. Note that in the corresponding
region, the oscillations in $C_q$ are largely suppressed, while $R_q$ deviates
from the universal value $e^2/(2h)$, which are attributed to the breaking of the
quantum coherence.\cite{gabelli2006, nigg2006}

\begin{figure}
  \centering
  \includegraphics[width=9cm]{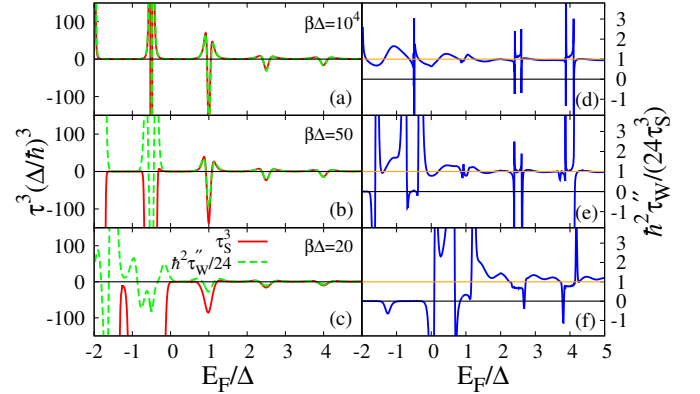}
  \caption{(Color online) The quantities $\tau^3_S$ (red solid curve) and
    $\hbar^2\tau^{''}_W/24$ (green dashed curve) as function of $E_F$ with
    inverse temperature $\beta \Delta = 10^4$ (a), $\beta \Delta = 50$ (b) and
    $\beta \Delta = 20$ (c). The thin black lines in (a-c) indicate the zero
    value. The corresponding ratios $\hbar^2\tau^{''}_W/(24\tau^3_S)$ are
    plotted in (d), (e) and (f), respectively. The thin black lines in (d-f)
    indicate the zero value, while the thin orange lines represent the value
    $1$. In all the figures, the charging energy $E_C$ is set to $0.5$.}
  \label{fig9}
\end{figure}

\begin{figure}
  \centering
  \includegraphics[width=8.5cm]{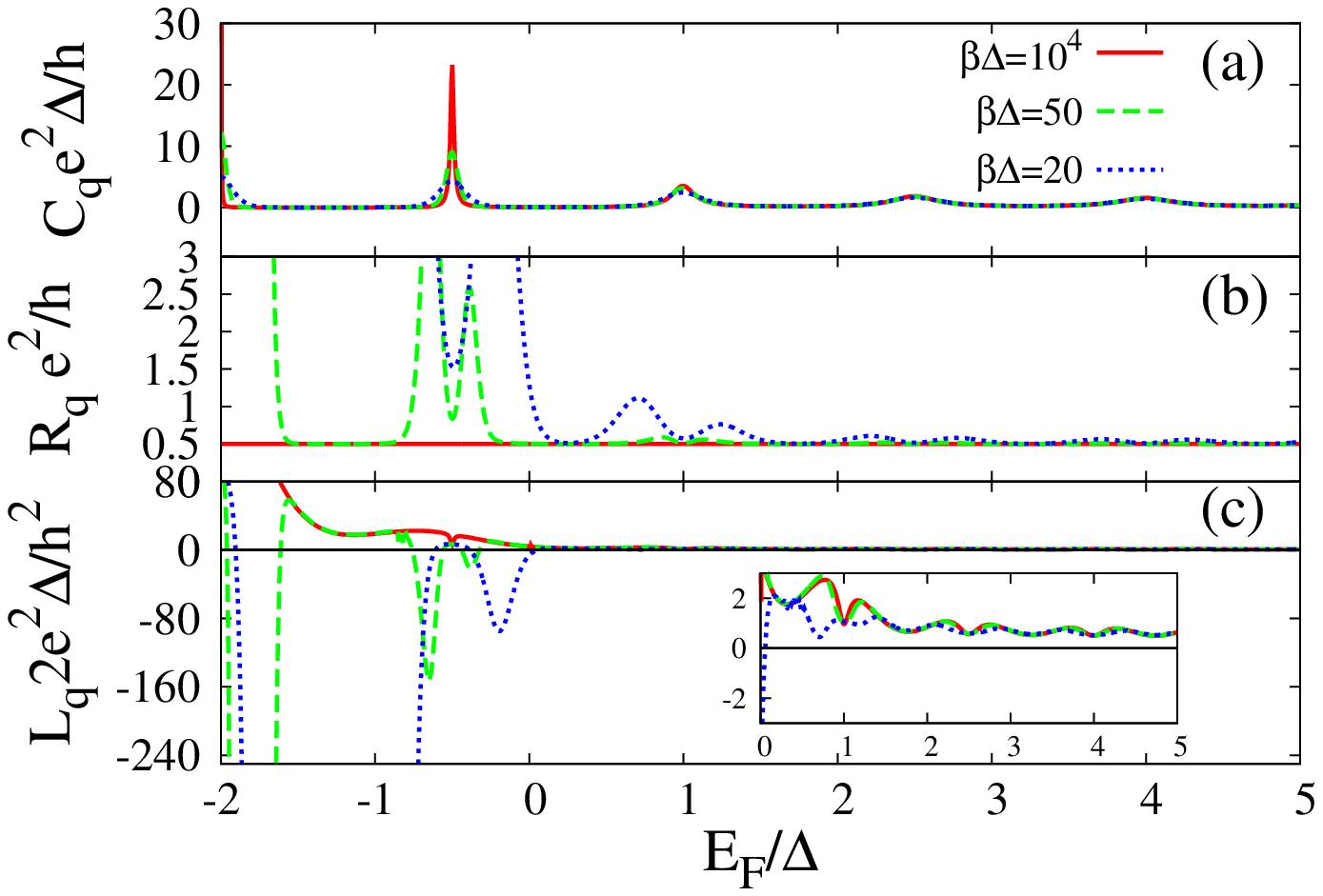}
  \caption{(Color online) The quantum capacitance $C_q$ (a) , charge relation
    resistance $R_q$ (b) and quantum inductance $L_q$ (c) as function of Fermi
    energy $E_F$ at different temperatures. The zooms of $L_q$ in the region
    $E_F/\Delta \in [0, 5]$ are plotted in the insets of (c).  The thin black
    lines in (c) and the inset indicate the zero value. In all the figures, the
    charging energy $E_C$ is set to $0.5$.}
  \label{fig10}
\end{figure}

\begin{figure}
  \centering
  \includegraphics[width=8.5cm]{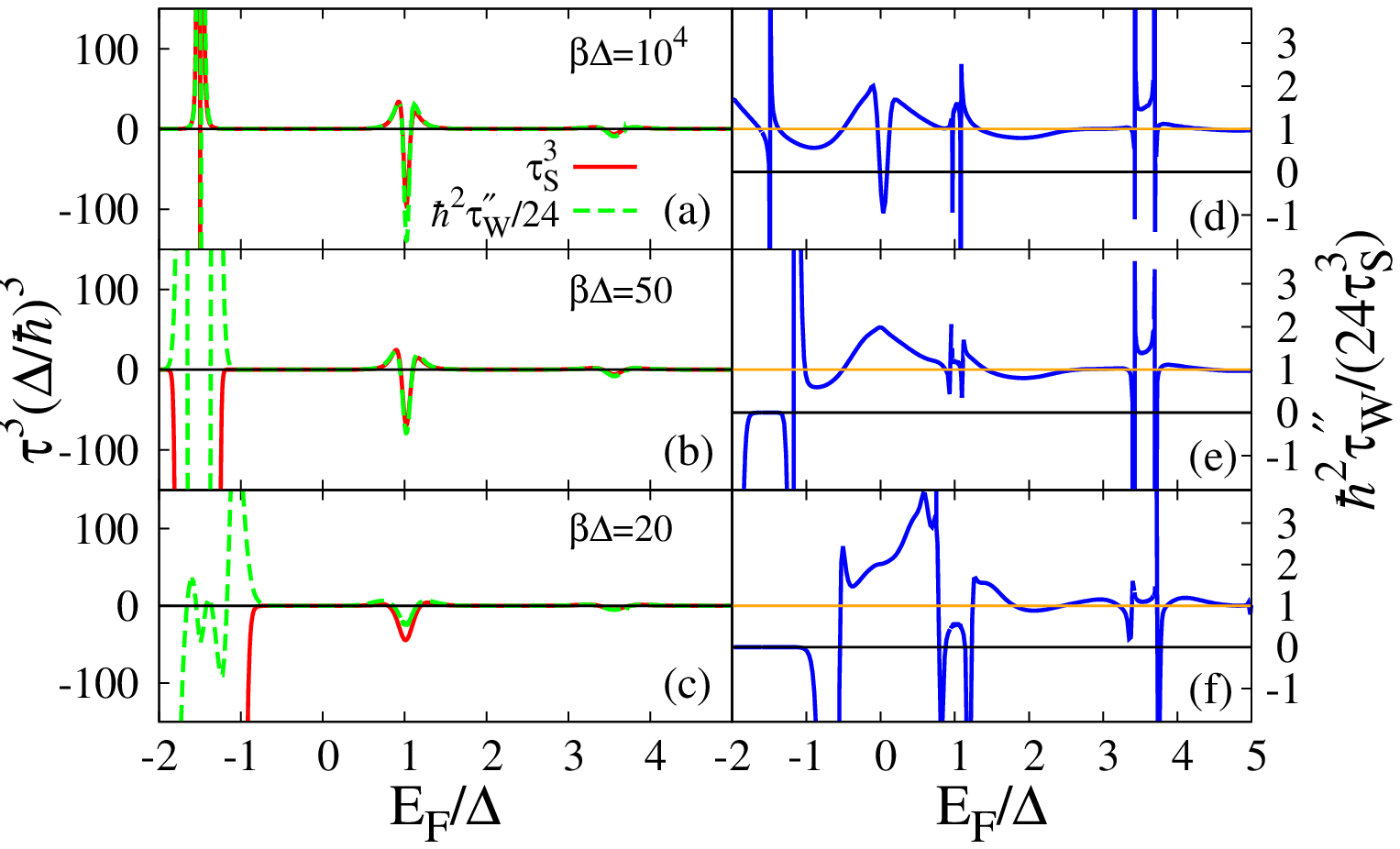}
  \caption{(Color online) The quantities $\tau^3_S$ (red solid curve) and
    $\hbar^2\tau^{''}_W/24$ (green dashed curve) as function of $E_F$ with
    inverse temperature $\beta \Delta = 10^4$ (a), $\beta \Delta = 50$ (b) and
    $\beta \Delta = 20$ (c). The thin black lines in (a-c) indicate the zero
    value. The corresponding ratios $\hbar^2\tau^{''}_W/(24\tau^3_S)$ are
    plotted in (d), (e) and (f), respectively. The thin black lines in (d-f)
    indicate the zero value, while the thin orange lines represent the value
    $1$. In all the figures, the charging energy $E_C$ is set to $1.5$.}
  \label{fig11}
\end{figure}

\begin{figure}
  \centering
  \includegraphics[width=8.5cm]{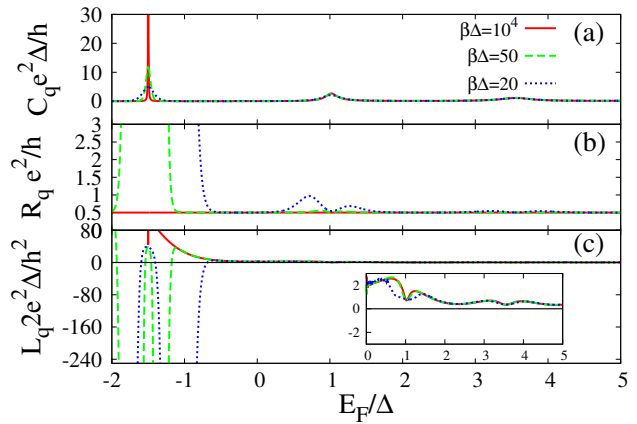}
  \caption{(Color online) The quantum capacitance $C_q$ (a) , charge relation
    resistance $R_q$ (b) and quantum inductance $L_q$ (c) as function of Fermi
    energy $E_F$ at different temperatures. The zooms of $L_q$ in the region
    $E_F/\Delta \in [0, 5]$ are plotted in the insets of (c).  The thin black
    lines in (c) and the inset indicate the zero value. In all the figures, the
    charging energy $E_C$ is set to $1.5$.}
  \label{fig12}
\end{figure}

In the presence of charging energy $E_C$, one can also find similar high
temperature behaviors of $\tau_S$ and $L_q$, which can be seen by comparing
Fig.~\ref{fig7}[Fig.~\ref{fig8}] to Fig.~\ref{fig9}[Fig.~\ref{fig10}] for
$E_C/\Delta=0.5$ and to Fig.~\ref{fig11}[Fig.~\ref{fig12}] for
$E_C/\Delta=1.5$. Note that as $E_C$ increases, the effect of the nonzero
temperature is less and less pronounced.

According to the previous discussion, one can conclude that as the temperature
increases, the ratio $\hbar^2\tau^{''}_W/(24\tau^3_S)$ goes from the value $1$
at low temperatures to $0$ at high temperatures. Accordingly, the quantum
inductance $L_q$ also show different behavior at high temperatures: It can go
from positive values at low temperatures to large negative values at high
temperatures. Hence, the relation $\hbar^2\tau^{''}_W/(24\tau^3_S) \approx 1$
and the positive definiteness of $L_q$ can be regarded as signatures of the ac
quantum coherent transport.

\section{SUMMARY}
\label{sec5}

In this work, we have examined the characteristic times which describe the low
frequency dynamics of the mesoscopic capacitors. By combining the
delayed-deformed current picture and the numerical calculations within NEGF
formalism, we found that the Wigner-Smith delay time $\tau_W$ can only capture
the ac response of the mesoscopic capacitor up to the second order of the
frequency.  At higher frequencies, a new time scale ---the deformation time
$\tau_S$--- has to be taken into consideration. The deformation time indicates
that due to the scattering, the profile of the outgoing current from the dot is
distorted from the incoming one. At sufficient low temperatures when the charge
transport is phase-coherent, $\tau_S$ can be approximated from the Wigner-Smith
delay time $\tau_W$ via a simple relation $\tau''_W/\tau^3_S=24/\hbar^2$. At
high temperatures when the coherence is broken, one has instead
$\tau''_W/\tau^3_S \to 0$. Hence this relation can be regarded as a signature of
the ac quantum coherent transport. We further show that the deformation time
$\tau_S$ can have a pronounced influence on the quantum inductance $L_q$ of the
mesoscopic capacitor, making $L_q$ show quite different behaviors from the ones
of the quantum capacitor $C_q$. The most striking one is that $L_q$ can change
its sign as the temperature increases: It goes from positive values at low
temperatures to large negative values at high temperatures. Thus the positive
definiteness of $L_q$ can also be regarded as a signature of the ac quantum
coherent transport. These results highlight the importance of the deformation
time on the ac response of the mesoscopic capacitors.

\begin{acknowledgments}
  The author would like to thank Professor J. Gao for bringing the problem to
  the author's attention. The author would also like to thank Professor
  D. S\'anchez for helpful discussion and comments. This work was supported by
  Key Program of National Natural Science Foundation of China under Grant
  No. 11234009 and National Key Technology R\&D Program of China under Grant
  No. 20-1125ZCKF.
\end{acknowledgments}

\appendix*

\section{Derivation of Eq.~\eqref{s3:eq7-2}}
\label{app}

We start from the expression of the quantum conductance\cite{buttiker1993,
  buttiker1993-1, buttiker1994}
\begin{eqnarray}
  g_q(\omega) & = & \frac{e^2}{h} \int dE [ 1 -
  S^{\dagger}(E-\frac{\hbar\omega}{2})S(E+\frac{\hbar\omega}{2}) ] \nonumber\\
  && \hspace{1cm} \times \frac{
    f(E-\frac{\hbar\omega}{2}) - f(E+\frac{\hbar\omega}{2}) }{\hbar \omega},
  \label{a1:eq1}
\end{eqnarray}
where $S(E) = e^{i \phi(E)}$ [Eq.~\eqref{s3:eq3-2}]. By performing a Taylor
expansion with respect to $\omega$, one obtains
\begin{eqnarray}
  g_q(\omega) & = & \frac{e^2}{h} \int dE [ -f'(E) -
  (\frac{\hbar\omega}{2})^2\frac{f'''(E)}{6} ] \nonumber\\
  && + \frac{e^2}{h} \int dE f'(E) e^{i \omega \tau_W(E) + i \omega^3 \tau^3_S(E) } \nonumber\\
  && + \frac{e^2}{h} \int dE (\frac{\hbar\omega}{2})^2\frac{f'''(E)}{6} e^{i
    \omega \tau_W(E) } \nonumber\\
  && + O(\omega^3).
  \label{a1:eq2}
\end{eqnarray}
At sufficient low temperatures, the Fermi distribution $f(E)$ can be well
approximated by the step function $\theta(E_F-E)$. By perform the integration
over $E$, one has
\begin{eqnarray}
  g_q(\omega) & = & \frac{e^2}{h} \nonumber\\
  &&{} - \frac{e^2}{h} e^{i \omega \tau_W(E_F) + i \omega^3 \tau^3_S(E_F) }
  \nonumber\\
  &&{} - \frac{e^2}{h} i \omega^3 \frac{\hbar^2\tau''_W(E_F)}{24} e^{i \omega
    \tau_W(E_F) } \nonumber\\
  &&{} + O(\omega^3).
  \label{a1:eq3}
\end{eqnarray}
By using the relation Eq.~\eqref{s3:eq6-3}, up to the second order of the
frequency $\omega$, the above equation can be approximated as
\begin{eqnarray}
  g_q(\omega) & \approx & \frac{e^2}{h} [1 - 2 e^{i \omega \tau_W(E_F) } e^{i \omega^3
    \tau^3_S(E_F)} \nonumber\\
  &&\hspace{2.5cm} + e^{i \omega \tau_W(E_F) }], \label{a1:eq4}
\end{eqnarray}
which is just the result given in Eq.~\eqref{s3:eq7-2}.


\begin{thebibliography}{0}
\bibitem{comment1} In the numerical calculation, we choose $\beta\Delta=10^4$ to
  describe the zero-temperature limit.
\end{thebibliography}
\end{document}